\documentclass[12pt]{article}

\usepackage{graphics,amssymb,epsfig,float}
\usepackage[usenames,dvips]{color}
\usepackage{graphicx}
\usepackage{epsfig}
\usepackage{rotating}
\usepackage{dcolumn}
\usepackage{bm}
\usepackage{cite}
\usepackage{amsmath}
\usepackage{indentfirst} 

\def\lsim{\;\raise0.3ex\hbox{$<$\kern-0.75em\raise-1.1ex\hbox{$\sim$}}\;}
\def\gsim{\;\raise0.3ex\hbox{$>$\kern-0.75em\raise-1.1ex\hbox{$\sim$}}\;}
\def\beq{\begin{equation}}   \def\eeq{\end{equation}}
\def\ba{\begin{array}}       \def\ea{\end{array}}
\def\bea{\begin{eqnarray}}   \def\eea{\end{eqnarray}}
\def\nn{\nonumber}

\def\k{\kappa}
\def\l{\lambda}

\def\ni{\noindent}

\def\CD{c_D}
\def\CV{c_V}

\def\CUi{c_{U_i}}
\def\CDi{c_{D_i}}
\def\CVi{c_{V_i}}
\def\CGi{c_{g_i}}
\def\CPi{c_{\gamma_i}}

\begin{document}
\begin{titlepage}
\begin{center}

\begin{flushright}
LAPTH-041/12\\
LPT Orsay 12-91
\end{flushright}

\vspace{1cm}
{\Large\bf Two Higgs Bosons at the Tevatron and the LHC?} 

\vspace{1cm}\renewcommand{\thefootnote}{\fnsymbol{footnote}}

{\large 
G.~Belanger$^{1}$\footnote[1]{Email: belanger@lapp.in2p3.fr},
U.~Ellwanger$^{2}$\footnote[2]{Email: Ulrich.Ellwanger@th.u-psud.fr},
J.~F.~Gunion$^{3}$\footnote[3]{Email: jfgunion@ucdavis.edu}, 
Y.~Jiang$^{3}$\footnote[4]{Email: yunjiang@ucdavis.edu},
S.~Kraml$^{4}$\footnote[5]{Email: sabine.kraml@lpsc.in2p3.fr}
} 

\renewcommand{\thefootnote}{\arabic{footnote}}

\vspace{1cm} 
{\normalsize \it 
$^1\,$LAPTH, Universit\'e de Savoie, CNRS, B.P.110, F-74941 Annecy-le-Vieux Cedex, France \\[1mm]
$^2\,$Laboratoire de Physique Th\'eorique, UMR 8627, CNRS and
Universit\'e de Paris--Sud, F-91405 Orsay, France\\[1mm]
$^3\,$Department of Physics, University of California, Davis, CA 95616, USA\\[1mm]
$^4\,$Laboratoire de Physique Subatomique et de Cosmologie, UJF Grenoble 1,
CNRS/IN2P3, INPG, 53 Avenue des Martyrs, F-38026 Grenoble, France
}

\vspace{1cm}

\begin{abstract}
The best fit to the Tevatron results in the $b\bar{b}$ channel and the
mild excesses at CMS in the $\gamma\gamma$ channel at 136~GeV and in the
$\tau\tau$ channel above 132~GeV can be explained by a second Higgs
state in this mass range, 
in addition to the one at 125~GeV recently discovered at the LHC.
We show that a scenario with two Higgs
bosons at 125~GeV and 136~GeV can be consistent with practically all available
signal rates, including a reduced rate in the $\tau\tau$ channel  around
125~GeV as reported by CMS. 
An example in the parameter space of the general NMSSM is given where, 
moreover, the signal rates of the 125~GeV Higgs boson in the 
$\gamma\gamma$ channels are enhanced 
relative to the expectation for a SM Higgs boson of this mass.
\end{abstract}

\end{center}

\end{titlepage}

\section{Introduction}

In July 2012 the ATLAS and CMS collaborations at the LHC announced
the observation of a new particle with the properties of a Higgs boson
\cite{:2012gk,:2012gu,ATLASWW}, a particle predicted in the Standard
Model of particle physics (SM). This observation was supported by
evidence for a Higgs boson found by the CDF and D0 collaborations at the
Tevatron \cite{:2012zzl}.

The SM makes quite precise predictions for the production cross sections
of the Higgs boson $H$ (via gluon-gluon fusion ($ggF$), vector boson
fusion ($VBF$) and associated production with an electroweak gauge
boson $V=W,Z$ ($VH$)), and its decay branching fractions into various
final states ($ZZ^{(*)}$, $WW^{(*)}$, $bb$, $\gamma \gamma$ and
$\tau \tau$) as a function of its unpredicted mass $M_H$. The observation
of the Higgs boson at the LHC is based primarily on the $\gamma
\gamma$ \cite{ATLASgamgam,CMSgamgam}, $ZZ^{(*)}$ \cite{ATLASZZ,CMSZZ}
and $WW^{(*)}$ \cite{ATLASWW,CMSWW} decay modes. The Higgs boson mass,
$M_H$, is quite precisely measured to be in the 125--126~GeV range using
the high resolution  $\gamma \gamma$ and $ZZ^{(*)}$ final states. The
evidence for the Higgs boson at the Tevatron is based principally on the
$bb$ decay mode \cite{Aaltonen:2012qt}, the observed enhancements
being consistent with a large range of possible Higgs masses.

Although most of the results originating from the various production and
decay channels are consistent with the expected properties of a SM Higgs
boson, differences of up to about two standard deviations are
observed in some cases:
\begin{enumerate}
\item the observed signal rate for $M_H=125$ GeV in the $\gamma \gamma$ decay mode is
enhanced relative to the SM expectation in the measurements of both the ATLAS and CMS collaborations
\cite{ATLASgamgam,CMSgamgam};
\item in the $\gamma \gamma$ decay mode, CMS observes an additional
excess corresponding to $M_H \sim 136$~GeV \cite{CMSgamgam};
\item the best fit to the measurement of $M_H$ at the Tevatron
corresponds to $M_H \sim 135$~GeV with an enhanced signal rate in the
$bb$ decay mode \cite{:2012zzl,Aaltonen:2012qt};
\item in the $\tau\tau$ decay mode, CMS observes a deficit for $M_H \sim
125$~GeV for the $VBF$-tag class of events, but an
excess for $M_H \gsim 132$~GeV~\cite{CMStautau}. \end{enumerate}

At present, these deviations are not significant enough to prove that
the Higgs sector is non-SM-like. However, in this paper we argue that
all of them can be described simultaneously by an extended Higgs sector
involving (at least) two states, one near 125~GeV \emph{and} a second near 136~GeV. We show
that the desired properties of these two Higgs bosons can be obtained in
the Next-to-Minimal Supersymmetric extension of the SM (NMSSM)
\cite{Ellwanger:2009dp}. In all models with an extended Higgs sector and
two light  CP-even Higgs bosons, the couplings of these Higgs bosons to
electroweak gauge bosons, up-type quarks, down-type quarks and charged
leptons, and the radiatively induced couplings to gluons and photons
will generally  differ from those of a SM Higgs boson; this makes it
possible to understand the discrepancies listed above with respect to
the SM.

In Section~2 we summarize the necessary properties of two Higgs bosons
such that these discrepancies are explained while maintaining consistency with the
numerous additional measurements of the various production and decay
channels by the ATLAS, CMS, CDF and D0 collaborations. In Section~3 we
briefly review the NMSSM, and discuss the properties of an appropriate
point in the NMSSM parameter space. Section~4 is devoted to conclusions.

\section{Desired properties of and constraints on two Higgs bosons}

For each production mode $\sigma(pp~\mbox{or}~ p\bar p\to Y \to
H_i)\equiv \sigma_{Y}(H_i)$ ($Y=gg,VV,VH_i$) and decay channel $H_i \to
X$ ($i=1,2$ or 3) of a non-standard Higgs boson, $H_i$, it is useful to
define a reduced signal rate $R_i^X(Y)$ relative to the expected signal
rate of a SM Higgs boson $H_{\rm SM}$ of the same mass:
\beq\label{eq:1}
R_i^X(Y) = \frac{\sigma_{Y}(H_i)}{\sigma_{Y}(H_{\rm SM})} \times
\frac{BR(H_i\to X)}{BR(H_{\rm SM}\to X)}
\eeq
For $Y=gg$ and $Y= VV$ we use the notations $ggF$ and $VBF$, respectively.
The first factor in \eqref{eq:1} we term the reduced production cross
section, the second factor in \eqref{eq:1} the reduced branching
fraction.

In addition, it is useful to define ``reduced couplings" (relative to the couplings of the SM Higgs boson) of the non-standard
Higgs bosons $H_i$ to electroweak gauge bosons ($\CVi$), to up-type
quarks ($\CUi$), to down-type quarks ($\CDi$), to gluons ($\CGi$) and to
photons ($\CPi$). We will make the assumption that the reduced couplings
to charged leptons such as the $\tau$-lepton are also given by $\CDi$,
as is the case in supersymmetric extensions of the SM
(up to radiative corrections relevant for very large $\tan\beta$).
If only
SU(2) doublet and singlet Higgs bosons exist, the $\CVi$ can never
exceed 1, and are identical for $V=W$ and $V=Z$. The radiatively
induced coupling to gluons originates mostly from top quark loops;
correspondingly one has $\CGi \sim \CUi$ if there are no additional SM-like or non-SM-like loop
contributions. The radiatively induced coupling to
photons originates mostly from the $W$ loop which leads to $\CPi \approx
\CVi$ neglecting the much smaller top loop contribution and loops involving non-SM particles.
(Note that one must not confuse $\CPi$ with the reduced
branching fraction into $\gamma\gamma$, see below.) 

In the case of the observed Higgs boson with a mass of 125--126 GeV, the
experimental collaborations have published best fits to reduced signal
rates for various decay modes. Sometimes these are combinations of
different production modes, in particular  $ggF$ and $VBF$, which
have to be disentangled. For all other Higgs masses, the experimental
collaborations have published bounds on (or p-values for) reduced signal
rates; again these correspond often to combinations of different
production modes. Using the available information, we will estimate the
production-mode specific signal rates \eqref{eq:1} for both a Higgs
boson $H_1$ with a mass of 125--126 GeV and a Higgs boson $H_2$ with a
mass of 135--136 GeV. From the production-mode specific signal rates we
can deduce the desired ranges of reduced couplings of both Higgs states.
We now sketch the corresponding steps.

The dependence of the reduced production cross sections of non-standard
Higgs bosons $H_i$ on the reduced couplings is given by (including most
radiative corrections)
\beq\label{eq:2}
\frac{\sigma_{i}( ggF)}{\sigma_{H_{\rm SM}}(ggF)}=(\CGi)^2\; ,\quad
\frac{\sigma_i({VBF})}{\sigma_{H_{\rm SM}}({VBF})}=
\frac{\sigma_i({VH})}{\sigma_{H_{\rm SM}}(VH)}=(\CVi)^2\; .
\eeq
However, the dependence of the reduced decay branching fractions of a
non-standard Higgs boson $H_i$ on the reduced couplings is more
involved. First, the partial widths $\Gamma(H_i \to X)$ have to be
rescaled correspondingly:
\bea\label{eq:3}
\Gamma(H_i \to bb) &=& (\CDi)^2\times \Gamma(H_{\rm SM} \to bb)\,, \nn \\
\Gamma(H_i \to WW^{(*)}) &=& (\CVi)^2\,\times \Gamma(H_{\rm SM} \to WW^{(*)})\,,  \\
\Gamma(H_i \to \gamma\gamma) &=& (\CPi)^2\,\times \Gamma(H_{\rm SM} \to \gamma\gamma),
\quad \text{etc.} \nn
\eea
Hence the total width is
\beq\label{eq:4}
\Gamma_{\rm Tot}(H_i) = (\CDi)^2\times \Gamma(H_{\rm SM} \to bb)
+ (\CVi)^2\times \Gamma(H_{\rm SM} \to WW^{(*)}) + \ \dots
\eeq
where we have shown only the dominant contributions. Finally the reduced
branching fractions are
\beq\label{eq:5}
\frac{BR(H_i\to X)}{BR(H_{\rm SM}\to X)} = \frac{\Gamma(H_i\to
X)}{\Gamma(H_{\rm SM}\to X)} \times \frac{\Gamma_{\rm Tot}(H_{\rm
SM})}{\Gamma_{\rm Tot}(H_i)}\; .
\eeq
The reduced signal rates of Eq.~\eqref{eq:1} can then be computed in
terms of the reduced couplings, Eqs.~\eqref{eq:2} and \eqref{eq:5}.

If two Higgs bosons with masses of about 125 GeV and 136 GeV exist, one
must distinguish Higgs search channels with high mass resolution for
which it is possible to measure the masses and reduced signal rates
separately (the $\gamma \gamma$ and $ZZ$ modes) from Higgs search
channels with low mass resolution to which the contributions from the
two states can overlap (the $WW$, $bb$ and $\tau\tau$ modes). Let
us begin with the present situation in the high mass resolution
channels. 

\vspace*{3mm}
\ni {\bf\boldmath 1) A Higgs boson $H_1$ at 125--126 GeV}

\vspace*{3mm}
The dominant production process allowing for the observation of a Higgs
boson in the $\gamma \gamma$ mode at the LHC is $ggF$. The best fits to
$R_1^{\gamma\gamma}(ggF)$ from the ATLAS and CMS collaborations are both
significantly larger than 1:  $R_1^{\gamma\gamma}(ggF)\simeq 1.8 \pm
0.5$ (ATLAS \cite{ATLASgamgam}), $R_1^{\gamma\gamma}(ggF)\simeq 1.5 \pm
0.5$ (CMS \cite{CMSgamgam}). Combining, we estimate
\beq\label{eq:6}
R_1^{\gamma\gamma}(ggF)\simeq 1.66 \pm 0.36\, .
\eeq
In addition, both collaborations have studied events with two additional
forward jets to which the contribution from $VBF$ is dominant.
At CMS, combining the 7~TeV dijet tag and 8~TeV dijet tight results, 
yields  $R_1^{\gamma\gamma}({VBF})\sim 2.6\pm 1.3$~\cite{CMSgamgam}; 
note that here the $VBF$ category contains roughly 25\% $ggF$ production.
For ATLAS, we obtain $R_1^{\gamma\gamma}({VBF})\sim 2.7\pm 1.5$~\cite{ATLASgamgam} (with unspecified $ggF$ contamination). Subsequently we merely assume $R_1^{\gamma\gamma}({VBF}) > 1$.

The $ggF$ process also dominates the Higgs signal in the $ZZ$ channel.
The best fits to $R_1^{ZZ^{(*)}}(ggF)$ from the ATLAS and CMS
collaborations are both consistent with 1:  $R_1^{ZZ^{(*)}}(ggF)\simeq
1.4 \pm 0.6$ (ATLAS \cite{ATLASgamgam}), $R_1^{ZZ^{(*)}}(ggF)\simeq 0.75
\pm 0.5$ (CMS \cite{CMSgamgam}). Combining, we estimate
\beq\label{eq:7}
R_1^{ZZ^{(*)}}(ggF)\simeq 1.02 \pm 0.38\, .
\eeq

\vspace*{3mm}
\ni {\bf\boldmath 2) A Higgs boson $H_2$ at 135--136 GeV}

\vspace*{3mm}

In the $\gamma \gamma$ mode, CMS has observed an excess of events of
about two standard deviations around 136~GeV \cite{:2012gu,CMSgamgam}, 
the excess being a bit larger for the 7 TeV data than for the 8~TeV data.
Combining the two data sets, the corresponding  reduced signal rate can be estimated as
$R_2^{\gamma\gamma}(ggF)\simeq 0.9 \pm 0.4$ (CMS). However, no excess of
events at this mass was observed by ATLAS \cite{ATLASZZ}. We estimate
$R_2^{\gamma\gamma}(ggF)\simeq 0.0 + 0.4$ (ATLAS). 
Taken together, one obtains
\beq\label{eq:8}
R_2^{\gamma\gamma}(ggF)\simeq 0.45 \pm 0.3 \, .
\eeq 
The above  is a crude estimate, which could be improved by more detailed
analyses and/or more data; here we consider it as a first hint for the
existence of a second state near 136~GeV in the Higgs sector.

In the $ZZ^{(*)}$ mode, no excess has been observed by the ATLAS and CMS
collaborations for $M_H \sim 136$~GeV \cite{ATLASZZ,CMSZZ}. Combining
both upper bounds on the reduced signal rate, we estimate
\beq\label{eq:9}
R_2^{ZZ^{(*)}}(ggF)\lsim 0.2
\eeq 
at the level of one standard deviation.

Next we turn to the low mass resolution channels.
In the $\tau\tau$ channel and tagging two jets (sensitive mostly to the
$VBF$ production mode), CMS observes a deficit with respect to the
background-only hypothesis assuming $M_H \sim 125$~GeV
\cite{:2012gu,CMStautau}. Hence $R_1^{\tau\tau}({VBF})$ should be as
small as possible.\footnote{In the $\tau\tau + (0/1)\,{\rm jets}$
channel, CMS observes a slight excess. The error bar is fairly large,
however, and taken together the 0/1 jets and $VBF$-tag channels still give
a deficit in the $\tau\tau$ channel.  
The ATLAS result for $H\to\tau\tau$ \cite{Aad:2012ur} is based on 7~TeV data only 
and excludes only about (4--5)$\times$ the SM rate; 
it is thus inconclusive for our purpose.}  
Assuming $M_H \gsim 132$~GeV, CMS observes an excess of events \cite{CMStautau} of 
about half a standard deviation; 
the upper limit on $R_2^{\tau\tau}(VBF)$ (for $M_{H_2} \sim 135$~GeV) is given as
\beq\label{eq:10}
R_2^{\tau\tau}({ VBF}) < 1.81\,.
\eeq
In the presence of two Higgs states these values have to be
reinterpreted. In principle, a state at $M_{H_2} \sim 135$~GeV can
contribute to the signal rate obtained assuming $M_{H_{\rm SM}} \sim
125$~GeV. However, the production cross section for $M_{H} \sim 135$~GeV
is about $30\%$ lower than for $M_{H} \sim 125$~GeV and, moreover, the
mass resolution is not very well known (estimated as $15\mbox{--}20\%$
in \cite{CMStautau}). Subsequently we assume that the contribution to
the signal rate obtained assuming $M_{H_{\rm SM}} \sim 125$~GeV from a
state at $M_{H_2} \sim 135$~GeV is not very large, without being able to
quantify it more precisely. Conversely, the contribution to the signal
rate obtained assuming $M_{H_{\rm SM}} \sim 135$~GeV from a state at
$M_{H_1} \sim 125$~GeV will not be large if $R_1^{\tau\tau}({ VBF})$
is small; in any case such a contribution can be tolerated given the
weak bound~\eqref{eq:10}.

In the $bb$ channel, the CDF and D0 collaborations at the Tevatron
(where the dominant production mode is $VH$) have observed large values
of $R^{bb}(VH)$: $R^{bb}_{125}(VH)\simeq 1.97+0.74-0.68$
assuming $M_{H_{\rm SM}}=125$~GeV, and
\beq\label{eq:11}
R^{bb}_{135}(VH)\simeq 3.53+1.26-1.16
\eeq
assuming $M_{H_{\rm SM}}=135$~GeV \cite{:2012zzl,Aaltonen:2012qt}. CMS
has also observed excesses in this channel \cite{:2012gu,CMSbb}, but
below the expectations for a SM Higgs boson at 125~GeV. Assuming larger
values of $M_{H_{\rm SM}}$, the excesses observed by CMS are larger
(with a peak around $M_{H_{\rm SM}} \sim 130$~GeV), but have large error
bars. 

It is clear that the central value of \eqref{eq:11} is difficult
to explain: the $VH$ production cross section $\propto \CV^2$ cannot be
enhanced with respect to the SM, and the SM Higgs branching fraction of
$\sim 40\%$ for $M_{H_{\rm SM}}=135$~GeV 
can be enhanced at most by a factor of 2.5 in the unphysical limit
$\CD \to \infty$.

Using the second of Eqs.~\eqref{eq:2} and the same reduced couplings of
Higgs bosons to $b$-quarks and $\tau$-leptons, one finds
\beq\label{eq:12}
R^{bb}(VH)=R^{\tau\tau}({ VBF})
\eeq
for all Higgs states. If $R_1^{\tau\tau}({VBF})$ is as small as
observed by CMS, the values for $R^{bb}(VH)$ measured at the
Tevatron should originate primarily from $H_2$ with
$M_{H_{2}}\sim$~135--136~GeV; this possibility is one of the main
advantages of the present proposal. However, the contribution of $H_1$
to the signal rate $R^{bb}(VH)$ obtained assuming $M_{H_{\rm SM}}
\sim 135$~GeV can still be sizable, since the production cross section
of $H_1$ is $\sim 30\%$ larger. Assuming a mass resolution 
worse than 10~GeV, $R^{bb}_{135}(VH)$ 
in \eqref{eq:11} would correspond to
\beq\label{eq:13}
R_{\rm eff}^{bb}(VH)\simeq R_{2}^{bb}(VH) + 1.3\times
R_{1}^{bb}(VH)\; .
\eeq
(In addition, the contribution from $H_2$ to the signal rate
$R^{bb}_{125}(VH)$ should be as large as possible.)

In the $WW^{(*)}$ channel (with $VH$-tag), all collaborations have
observed excesses over a large mass range up to $M_H\sim 150$~GeV
\cite{:2012gk,:2012gu,ATLASWW,CMSWW}. Given the low mass resolution in
this channel and the correspondingly large error bars, the measured
values of $R^{WW^{(*)}}(VH)$ do not impose additional constraints on a
scenario with two Higgs bosons at 125--126 and 135--136 GeV.

What are the consequences of the above results on the reduced couplings
of the two Higgs bosons proposed here? First, the enhanced signal rate
$R_1^{\gamma\gamma}(ggF)$ at 125--126~GeV, Eq.~\eqref{eq:6}, has to be
explained. It has been observed in several publications
\cite{Carena:1999bh,Ellwanger:2010nf,Carena:2011aa,Hall:2011aa,Ellwanger:2011aa,King:2012is,Cao:2012fz,
Vasquez:2012hn,Ellwanger:2012ke,Jeong:2012ma,Blum:2012ii,Benbrik:2012rm,
Gunion:2012gc,SchmidtHoberg:2012yy}
that the branching
fraction of a non-standard Higgs boson into $\gamma\gamma$ is enhanced
if its coupling $\CD$ to down-type quarks is reduced --- a reduction of
$\CD$ reduces the (dominant) partial width into $bb$ and hence the
total width $\Gamma_{\rm Tot}$;  
in turn, a reduced $\Gamma_{\rm Tot}$ in the denominator of \eqref{eq:5}
will increase the (reduced) branching fraction into $\gamma\gamma$. 
(Furthermore the reduced coupling $c_{g_1}\sim c_{U_1}$ of
$H_1$ to gluons should not be small.)
If
$\CD$ coincides with the reduced coupling to $\tau$-leptons, a reduced
branching fraction of $H_1$ into $\tau\tau$ fits well with the small
value of $R_1^{\tau\tau}(VBF)$ observed by CMS. Of course, due to
\eqref{eq:12}, a reduced signal rate of $H_1$ into $bb$ in $VH$
would be in obvious conflict with the Tevatron results if no other Higgs
boson would exist. Thus, the reduced coupling $c_{D_2}$ of $H_2$ had better
be enhanced.

A reduced total width $\Gamma_{\rm Tot}$ of $H_1$ due to a reduced
partial width into $bb$ can also increase its branching fraction
into $ZZ$; together with a slight reduction of $c_{V_1}$ and $c_{g_1}$
the SM-like value of $R_1^{ZZ^{(*)}}(ggF)$ in
\eqref{eq:7} is a natural result.

For $H_2$, the coupling $c_{g_2}\sim c_{U_2}$ to gluons must be smaller than 1
in order to comply with \eqref{eq:8} and \eqref{eq:9}. However,
$c_{D_2}$ (and hence the branching fraction into $bb$) should be
enhanced, and $c_{V_2}$ should not be small in order to comply with
\eqref{eq:11} together with \eqref{eq:13}.

In the next section we will briefly discuss the NMSSM, and present a
point in the NMSSM parameter space where $H_1$ and $H_2$ have masses of $125$ GeV and $136$ GeV, respectively, and have the desired reduced couplings. (In the
CP-conserving MSSM with its two CP-even Higgs bosons we were not able to
find such points in the much more constrained parameter space.)

\section{Two Higgs bosons at 125 and 136 GeV in the NMSSM}

The NMSSM is the simplest supersymmetric (SUSY) extension of the SM with
a scale invariant superpotential, and does not suffer from the
$\mu$-problem of the MSSM (the presence of a SUSY mass parameter whose
value must accidentally be of order $M_{\rm SUSY}$, the mass scale of the soft SUSY
breaking terms). The Higgs sector of the NMSSM contains two doublet
superfields $H_u$ and $H_d$ (with couplings of $H_u$ to up-type quarks,
and couplings of $H_d$ to down-type quarks and leptons as in the
MSSM) and an additional SU(2)-singlet superfield~$S$. The
NMSSM-specific part of the superpotential is
\beq\label{eq:14}
W_{\rm NMSSM}=\l S H_u H_D + \frac{\k}{3}S^3\; ,
\eeq
where the first term generates an effective $\mu$-term with $\mu_{\rm
eff}=\l s$ once the scalar component of $S$ develops a vacuum expectation
value (vev) $s$. The vev $s$ is triggered by the NMSSM-specific soft SUSY
breaking terms (from here onwards, $H_u$, $H_d$ and $S$ denote the
scalar components of the corresponding superfields),
\beq\label{eq:15}
{\cal L}({\rm soft})_{\rm NMSSM}=-m_S^2 S^2-\l A_\l S H_u H_d -
\frac{\k}{3} A_\k S^3\; ,
\eeq
and is thus  naturally of order $M_{\rm SUSY}$. The field content in the
Higgs sector of the NMSSM consists of three neutral CP-even bosons $H_i$,
$i=1\dots 3$, two neutral CP-odd bosons $A_i$, $i=1\dots 2$, and a charged
Higgs boson $H^\pm$. 

The CP-even bosons $H_i$ are linear combinations of the real components
of $H_u$, $H_d$ and $S$. Their masses and mixing angles have to be
obtained from the $3\times 3$ mass matrix including SUSY terms, soft
SUSY breaking terms and radiative corrections. Expressions for the mass
matrices of the physical CP-even and CP-odd Higgs states---after $H_u$,
$H_d$ and $S$ have acquired vevs $v_u$, $v_d$ and $s$ and including the
dominant radiative corrections---can be found in \cite{Ellwanger:2009dp}
and will not be repeated here. The Higgs sector of the NMSSM is described
by the six parameters
\beq \label{eq:16}
\lambda\ , \ \kappa\ , \ A_{\lambda} \ , \ A_{\kappa}, \ \tan \beta\ =
v_u/v_d\ ,\ \mu_\mathrm{eff}\; .
\eeq

The couplings of the Higgs states depend on their decompositions into
the CP-even weak eigenstates $H_d$, $H_u$ and $S$, which are given by
\bea\label{eq:17}
H_1 &=& S_{1,d}\ H_d + S_{1,u}\ H_u +S_{1,s}\ S\; ,\nn \\
H_2 &=& S_{2,d}\ H_d + S_{2,u}\ H_u +S_{2,s}\ S\; .
\eea
Then, the reduced couplings of $H_i$ are
\beq\label{eq:18}
\CDi = \frac{S_{i,d}}{\cos\beta}\;,\qquad
\CUi = \frac{S_{i,u}}{\sin\beta}\;,\qquad
\CVi = \cos\beta\,S_{i,d} + \sin\beta\, S_{i,u}\; .
\eeq
The loop-induced reduced couplings $\CGi$ and $\CPi$ have to be computed
including contributions from SUSY particles in the loops, including
scalar $\tau$-leptons, charginos and more. Below, the
calculations of Higgs masses, mixing angles and reduced couplings have
been performed by the code NMSSMTools$\_$3.2.1~\cite{Ellwanger:2004xm,Ellwanger:2005dv} 
including radiative corrections
to the Higgs sector from~\cite{Degrassi:2009yq}.

Next we present a point in the parameter space of the general NMSSM with
the desired properties. For the MSSM-like soft SUSY breaking terms we
choose bino, wino and gluino masses $M_1$=220~GeV, $M_2$=400~GeV and
$M_3$=1100~GeV respectively, squark masses of 1500~GeV for the first two
generations and the right-handed $b$-squarks, 1000~GeV for sleptons and
the other third generation squarks, and finally $A_t = A_b = -2500$~GeV,
$A_\tau=-1000$~GeV.
The NMSSM-specific input parameters are listed in Table~1, together with
the resulting masses of the various Higgs states. 

\begin{table}
\begin{center}
\begin{tabular}{|c|c||c|c|} \hline
$\lambda$               & 0.617&$\mu_{\text{eff}}$   &  143      \\\hline
$\kappa$                & 0.253 & $A_\l$ & 164 \\\hline
$\tan\beta$             &  1.77& $A_\k$  & 337   \\\hline  \hline
$M_{H_1}$   & 125 & $M_{A_1}$ &95  \\\hline
$M_{H_2}$   & 136 & $M_{A_2}$ &282  \\\hline
$M_{H_3}$   & 289 & $M_{H^\pm}$& 272  \\\hline
\end{tabular}
\caption{NMSSM-specific parameters and Higgs masses of a point with
desired properties. (The dimensionful parameters are given in GeV.)}
\end{center}
\end{table}

\begin{table}
\begin{center}
\begin{tabular}{|c|c|c|c|c|c|c|c|c|} \hline
Higgs & $S_{i,d}$ & $S_{i,u}$ & $S_{i,s}$ & $\CDi$ & $\CUi$ & $\CVi$ & $\CGi$& $\CPi$
  \\\hline
$H_1$ & $-0.24$ & $-0.67$ & 0.70  & $-0.48$ & $-0.77$ & $-0.70$ & 0.77 & 0.85  \\\hline
$H_2$ &  0.54 &  0.51 & 0.67  & 1.09 & 0.58 &  0.71 & 0.54 & 0.66  \\\hline
$H_3$ &  0.81 & $-0.54$ & $-0.24$ & 1.64 & $-0.62$ & -0.07 & 0.65 & 0.28  \\\hline
\end{tabular}
\caption{Mixing parameters \eqref{eq:17} and reduced couplings of the
three CP-even Higgs states.}
\end{center}
\end{table}

The decompositions $S_{i,j}$ and the reduced couplings of the 3 CP-even
Higgs states are given in Table~2. We see that the Higgs states are
strongly mixed, both $H_1$ and $H_2$ having large SU(2) doublet and
singlet components. $H_1$ has the smallest $c_D$ component, which leads
to an increase of the reduced branching fraction into $\gamma\gamma$ as
discussed above. However, the partial width $\Gamma(H_1 \to \gamma\gamma)$ also 
receives an additional NMSSM-specific contribution
of $\sim 20\%$ from higgsino-like charginos with $m_{\tilde\chi^\pm_1}=126$~GeV in the loop;
this possibility was mentioned previously in~\cite{King:2012is,SchmidtHoberg:2012yy}.

Finally, we give the reduced branching fractions for 
the CP-even Higgs bosons in Table~3, and their signal rates relative to SM expectations in Table~4.

\begin{table}
\begin{center}
\begin{tabular}{|c|c|c|c|} \hline
Higgs& $\frac{BR(H_i\to bb)}{BR(H_{\rm SM}\to bb)}$ & 
$\frac{BR(H_i\to VV^{(*)})}{BR(H_{\rm SM}\to VV^{(*)})}$ & 
$\frac{BR(H_i\to \gamma\gamma)}{BR(H_{\rm SM}\to \gamma\gamma)}$   \\\hline
$H_1$ & 0.73 & 1.52 & 2.21    \\\hline
$H_2$ & 1.46 &  0.62 & 0.54   \\\hline
$H_3$ & 43.45 & 0.08 & 1.37   \\\hline
\end{tabular}
\caption{Reduced branching fractions for the three CP-even Higgs states.
Note that we have  $\frac{BR(H_i\to \tau\tau)}{BR(H_{\rm SM}\to
\tau\tau)} \sim \frac{BR(H_i\to bb)}{BR(H_{\rm SM}\to
bb)}$,  and $\frac{BR(H_i\to WW^{(*)})}{BR(H_{\rm SM}\to
WW^{(*)})} = \frac{BR(H_i\to ZZ^{(*)})}{BR(H_{\rm SM}\to ZZ^{(*)})}
\equiv \frac{BR(H_i\to VV^{(*)})}{BR(H_{\rm SM}\to VV^{(*)})}$.}
\end{center}
%
\begin{center}
\begin{tabular}{|c|c|c|c|c|c|c|c|c|} \hline
Higgs &
$R^{\gamma\gamma}(ggF)$ & $R^{\gamma\gamma}(VBF)$ & $R^{VV^{(*)}}(ggF)$ & 
$R^{VV^{(*)}}(VH)$& $R^{bb}(VH)$ & $R^{\tau\tau}(ggF)$  \\\hline
$H_1$ & 1.30 & 1.09 & 0.90  & 0.75 & 0.36 & 0.42  \\\hline
$H_2$ & 0.16 &  0.27 & 0.18  & 0.31 & 0.74 &  0.43    \\\hline
$H_3$ & 0.58 & 0.01 & 0.04 & 0.004 & 0.23 & 19.6   \\\hline
\end{tabular}
\caption{Reduced signal rates for the three CP-even Higgs states.
Note that \newline
$R^{VV^{(*)}}({VBF})=R^{VV^{(*)}}(VH)$, and $R^{\tau\tau}({VBF})\sim
R^{bb}(VH)$.}
\end{center}
\end{table}

Let us now examine the extent to which these signal rates have the desired
properties listed in Section~2.  We observe that  $R_1^{\gamma\gamma}(ggF)$,
$R_1^{ZZ^{(*)}}(ggF)$, $R_2^{\gamma\gamma}(ggF)$ and
$R_2^{ZZ^{(*)}}(ggF)$ satisfy Eqs. \eqref{eq:6}, \eqref{eq:7},
\eqref{eq:8} and \eqref{eq:9}, respectively. Note that 
$R_1^{\gamma\gamma}(VBF)$ is also enhanced, in agreement with the
observations. In the $\tau\tau$ channel,
$R_1^{\tau\tau}(VBF)=R_1^{bb}(VH)$ 
is indeed suppressed,  as is $R_1^{\tau\tau}(ggF)$.  
$R_2^{\tau\tau}({VBF})$ is not enhanced but, as discussed in
Section~2, $R_2^{\tau\tau}$ (like $R_2^{bb}(VH)$) can receive a
considerable contribution from $R_1^{\tau\tau}$.  For $R_{\rm
eff}^{bb}(VH)$ as defined in \eqref{eq:13} we obtain $R_{\rm
eff}^{bb}(VH)\sim 1.20$, with the dominant contribution from
$R_2^{bb}(VH)$. This value coincides with the large excess given
in \eqref{eq:11} (assuming a single Higgs state at 135~GeV) only within
about two standard deviations, but at least exceeds the SM value.
Finally, the signal rates in the $WW^{(*)}$ channel via $VH$ are
consistent with the present limits.

The third CP-even Higgs state $H_3$ with  mass of about 290~GeV has
properties similar to the heavy scalar Higgs $H$ in the MSSM, in that it has an
enhanced signal rate in the $gg \to H_3 \to bb/\tau\tau$ channels
and suppressed couplings to electroweak gauge bosons. However, due to
the low value of $\tan\beta$, which is typical
for NMSSM scenarios such as the one discussed here, the present constraints on such a
state from direct searches~\cite{Chatrchyan:2012vp,ATLAS-CONF-2012-094}  
as well as the $B$-physics constraints implemented in NMSSMTools$\_$3.2.0     
are well satisfied.

Finally we have also attempted to look for similar scenarios in the
Higgs sector of the semi-constrained NMSSM~\cite{Gunion:2012zd,Ellwanger:2012ke,Gunion:2012gc}, 
where one requires universal soft
SUSY breaking terms at the GUT scale except for the Higgs soft-SUSY-breaking mass terms and
the NMSSM-specific trilinear couplings $A_\l$ and $A_\k$. In fact, one can
find scenarios where $H_1$ and $H_2$ have masses of about 125 and
136~GeV, respectively. 
However, we did not find any points where the constraints \eqref{eq:6}
to \eqref{eq:9} are all satisfied simultaneously;  at least one of the
conditions on $R_1^{\gamma\gamma}(ggF)$, $R_2^{\gamma\gamma}(ggF)$ or 
$R_2^{ZZ^{(*)}}(ggF)$ has to be relaxed to find valid points. For
example, we can satisfy Eqs.~\eqref{eq:6}, \eqref{eq:7}, \eqref{eq:9}
(and \eqref{eq:10}), but then $R_2^{\gamma\gamma}(ggF)$ turns out too
low, $R_2^{\gamma\gamma}(ggF)\lesssim 0.06$. Or we can satisfy
\eqref{eq:7}--\eqref{eq:10}, but then $R_1^{\gamma\gamma}(ggF)\lesssim
1.3$.   Moreover, $R_2^{bb}(VH)$ is never large, making it
difficult to explain the Tevatron result in this channel.

\section{Conclusions}

In the present paper we propose that the best fit to the Tevatron
results in the $bb$ channel and to the mild excesses at CMS in the
$\gamma\gamma$ channel at 136~GeV and in the $\tau\tau$ channel above
132~GeV could point towards a second Higgs state with this mass. In the
NMSSM, where an enhanced signal rate of a 125~GeV Higgs boson in the
$\gamma\gamma$ channel can be explained by a strong mixing between the SU(2)
doublet and the singlet states, 
one indeed expects
at least one additional state with a mass not too far from
125~GeV. We have shown an example of such a point in the parameter
space of the general NMSSM. However, an enhanced signal rate in the
$VH\to bb$ coinciding with the central value of about 3.5 ($\pm
1.2$) as measured by the CDF and D0 collaborations at the Tevatron
is practically impossible to obtain even if one assumes that it is due
to a superposition of two distinct Higgs states; 
nevertheless, signal rates above the SM value for a 136~GeV Higgs boson
are possible in such a scenario.

The scenario proposed here can be verified by measurements in the
$\gamma\gamma$ and $ZZ$ channels with high mass resolution; however, the
sensitivities to the signal rates of a 136~GeV Higgs boson have to be
about 4--5 times better than for a SM Higgs boson of such mass. 
Further tests of the scenario involve searches for the heavy Higgs doublet-like $H_3$, 
which cannot be too heavy here since it participates in the mixing. 
Note however that $H_3$ and $A_2$ can have large
branching fractions into the lighter Higgs states and/or neutralino pairs so that
non-standard search channels need to be exploited. 

The NMSSM also predicts the presence of SUSY
particles like squarks, gluinos and charginos/neutralinos. The colored
sparticles should eventually be observed at the LHC, although their decay channels
are possibly more complicated than in the MSSM, see
e.g.~\cite{Das:2012rr,Vasquez:2012hn}. 
Furthermore, scenarios with an enhanced di-photon
rate for the Higgs are typically associated with
small values of $\mu_{\text{eff}}$ implying light higgsino-like charginos/neutralinos 
that can be accessible in direct production at the LHC.
Note, however, that the pure higgsino-LSP (lightest SUSY particle) case is extremely 
difficult to detect at the LHC~\cite{Baer:2011ec}.

\section*{Acknowledgements} 

This work has been supported in part by US DOE grant DE-FG03-91ER40674
and by IN2P3 under contract PICS FR--USA No.~5872.  
UE acknowledges partial support from the French ANR LFV-CPV-LHC, ANR
STR-COSMO and the European Union FP7 ITN INVISIBLES (Marie Curie
Actions,~PITN-GA-2011-289442).
GB, UE, JFG, and
SK acknowledge the hospitality and the inspiring working atmosphere   of
the Aspen Center for Physics which is supported by the National Science
Foundation Grant No.\ PHY-1066293.


\end{document}